\documentclass[11pt]{article}

\usepackage[final]{acl}
\usepackage{times}
\usepackage{latexsym}
\usepackage{booktabs}
\usepackage{multirow}
\usepackage{amsmath}
\usepackage[T1]{fontenc}
\usepackage[utf8]{inputenc}
\ifdefined\XeTeXversion
  \catcode"2019=\active
  \begingroup
    \lccode`\~="2019
    \lowercase{\endgroup\def~}{\textquoteright}
\else
  \DeclareUnicodeCharacter{2019}{\textquoteright}
\fi
\usepackage{microtype}
\usepackage{subcaption}
\usepackage{graphicx}

\makeatletter
\let\scope@thebibliography\thebibliography
\newif\ifscope@targetfirstbibitem
\newif\ifscope@targetsecondbibitem
\newif\ifscope@openbibblock
\newcount\scope@newblockcount
\def\scope@targetbibkeyfirst{manzoor-etal-2026-multi}
\def\scope@targetbibkeysecond{baly-etal-2020-written}
\def\scope@closebibblock{%
  \ifscope@openbibblock
    \par\end{samepage}%
    \scope@openbibblockfalse
  \fi
}
\def\scope@preparebibitem#1{%
  \scope@closebibblock
  \def\scope@thisbibkey{#1}%
  \scope@targetfirstbibitemfalse
  \scope@targetsecondbibitemfalse
  \ifx\scope@thisbibkey\scope@targetbibkeyfirst
    \scope@targetfirstbibitemtrue
  \fi
  \ifx\scope@thisbibkey\scope@targetbibkeysecond
    \scope@targetsecondbibitemtrue
  \fi
  \scope@newblockcount=0
}
\def\scope@bibitem{\@ifnextchar[\scope@bibitem@opt\scope@bibitem@plain}
\def\scope@bibitem@opt[#1]#2{%
  \scope@preparebibitem{#2}%
  \scope@originalbibitem[#1]{#2}%
}
\def\scope@bibitem@plain#1{%
  \scope@preparebibitem{#1}%
  \scope@originalbibitem{#1}%
}
\def\scope@newblock{%
  \advance\scope@newblockcount by 1
  \ifscope@targetfirstbibitem
    \ifnum\scope@newblockcount=1
      \par\begin{samepage}\scope@openbibblocktrue
    \else
      \ifnum\scope@newblockcount=2
        \scope@closebibblock
      \fi
    \fi
  \fi
  \ifscope@targetsecondbibitem
    \ifnum\scope@newblockcount=2
      \par\begin{samepage}\scope@openbibblocktrue
    \fi
  \fi
  \scope@originalnewblock
}
\renewcommand{\thebibliography}[1]{%
  \scope@thebibliography{#1}%
  \let\scope@originalbibitem\bibitem
  \let\scope@originalnewblock\newblock
  \let\scope@originalendthebibliography\endthebibliography
  \let\bibitem\scope@bibitem
  \let\newblock\scope@newblock
  \def\endthebibliography{%
    \scope@closebibblock
    \scope@originalendthebibliography
  }%
}
\makeatother

\title{SCOPE: Source Credibility and Outlet Profiling Engine}

\author{
  Ilyas Nagiyev$^{1}$\thanks{\ \,Equal contribution.}, \;
  Miras Baisbay$^{1}$\footnotemark[1], \;
  Daniil Orel$^{2}$\footnotemark[1], \;
  Preslav Nakov$^{2}$ \\[3pt]
  $^{1}$Nazarbayev University \qquad $^{2}$MBZUAI \\[3pt]
  \texttt{\{ilyas.nagiyev, miras.baisbay\}@nu.edu.kz} \\
  \texttt{\{daniil.orel, preslav.nakov\}@mbzuai.ac.ae}
}

\begin{document}
\raggedbottom
\maketitle
\begin{abstract}
We present SCOPE, a multi-agent system for profiling news media outlets with respect to bias, factuality, and credibility. Given an outlet URL, SCOPE coordinates specialized agents for article acquisition, content analysis, external evidence search, scoring, and report generation. The system follows an MBFC-style methodology while making intermediate evidence and agent outputs inspectable through a structured demonstration interface. On a benchmark of 200 media outlets, SCOPE outperforms LLM-only, article-only, search-only, and media-background-check baselines on most metrics, achieving the lowest bias and factuality MAE, perfect failed-fact-check detection, and the lowest factual error rate. These results suggest that decomposed, evidence-grounded agent workflows can support more transparent and reliable media profiling. Our code is available at \url{https://github.com/mbaisbay/SCOPE}, and a live demonstration at \url{https://scope-demo.streamlit.app/}. Our video demonstration is available via the \href{https://drive.google.com/file/d/1uaouiXnwjdWV5ESc0PdDPRiLyBWS3Tri/view?usp=sharing}{URL (clickable)}.
\end{abstract}

\section{Introduction}
The rapid expansion of online media has created an increasingly complex and fragmented information landscape, making it difficult for readers, researchers, journalists, and policy analysts to systematically evaluate the reliability, political leaning, and broader influence of news sources at scale.

While much work on misinformation focuses on verifying individual claims~\cite{xie-etal-2025-fire,wang-etal-2025-openfactcheck}, detecting misleading articles, or evaluating the manipulation techniques applied by media outlets~\cite{sajwani-etal-2024-frappe,da-san-martino-etal-2020-prta}, other work has emphasized outlet-level profiling, such as estimating factuality, political ideology, framing, stance, and propagandistic content~\cite{zhang-etal-2019-tanbih}.

Yet many real-world decisions require understanding the media outlet itself, including its topical focus, ownership, citation practices, audience positioning, and past behavior.

Media profiling is particularly valuable when claims are new, ambiguous, or hard to verify directly. In such cases, an outlet's long-term patterns offer useful context for interpreting its content.
Building reliable media profiles remains challenging because relevant evidence is distributed across many sources, including outlet articles, fact-checking platforms, social media accounts, public records, and independent third-party reports.

A single model may struggle to collect relevant evidence, evaluate its reliability, reconcile conflicting signals, and produce a transparent profile that users can inspect. Media profiling is not merely a classification task, but a multi-stage analytical process involving retrieval, extraction, comparison, reasoning, and explanation.

In this paper, we introduce SCOPE\footnote{\underline{S}ource \underline{C}redibility and \underline{O}utlet \underline{P}rofiling \underline{E}ngine}, a multi-agent media-profiling tool that distributes this workflow among specialized agents. Each agent performs a distinct function, including data acquisition, content analysis, evidence search, or report generation.
By coordinating these agents, SCOPE combines complementary evidence streams to produce structured and explainable media profiles.

This design aims to improve coverage, reduce the burden on any single model, and make the profiling process more transparent through intermediate evidence and agent-level outputs.

\section{System Description}
In this section we describe how SCOPE works, which parts are included in the system, and the rationale behind our design choices.

\subsection{Methodology}
Media profiling requires a systematic methodology, since assessments of bias and reliability may otherwise become subjective. We follow the methodology proposed by Media Bias/Fact Check (MBFC)\footnote{\url{https://mediabiasfactcheck.com/methodology/}}, an independent media-rating organization that evaluates news sources with respect to ideological bias, factual reporting, transparency, and credibility.

MBFC evaluates each outlet through a human review process that includes analysis of at least ten headlines and five full news articles. The review considers language, framing, tone, sourcing practices, balance of viewpoints, and consistency with verified external information. Additional factors, such as ownership, funding, authorship, and editorial transparency, are also considered when available.

The bias rating is computed as a weighted score on a left--right scale from $-10$ to $+10$. MBFC defines the overall bias score as
\[
B = 0.35E + 0.35S + 0.15N + 0.15O,
\]
where $E$ denotes the outlet's economic ideology, $S$ its position on social values, $N$ the balance of its straight news reporting, and $O$ its editorial or opinion bias. 
Economic and social dimensions receive the largest weights because they capture the main ideological orientation of a media outlet. Straight news balance and editorial bias are included, since bias may also appear through story selection, framing, omission of opposing views, or emotionally loaded language. Negative values of $B$ indicate left-leaning bias, positive values indicate right-leaning bias, and values close to zero correspond to least-biased reporting.


Factual reliability is computed from failed fact checks ($C$), sourcing quality ($R$), transparency ($T$), and one-sidedness, omission, or propaganda ($M$):
\[
F = 0.40C + 0.25R + 0.25T + 0.10M.
\]
Failed fact checks receive the highest weight as the strongest indicator of unreliability. Sourcing and transparency measure whether outlets provide credible evidence and disclose their ownership, funding, authorship, and editorial responsibility. The smaller weight for $M$ reflects its overlap with bias while accounting for selective reporting and omission. Lower values of $F$ indicate higher factual reliability.

\begin{figure*}[t]
    \centering
    \begin{subfigure}{0.325\linewidth}
        \centering
        \includegraphics[width=\linewidth]{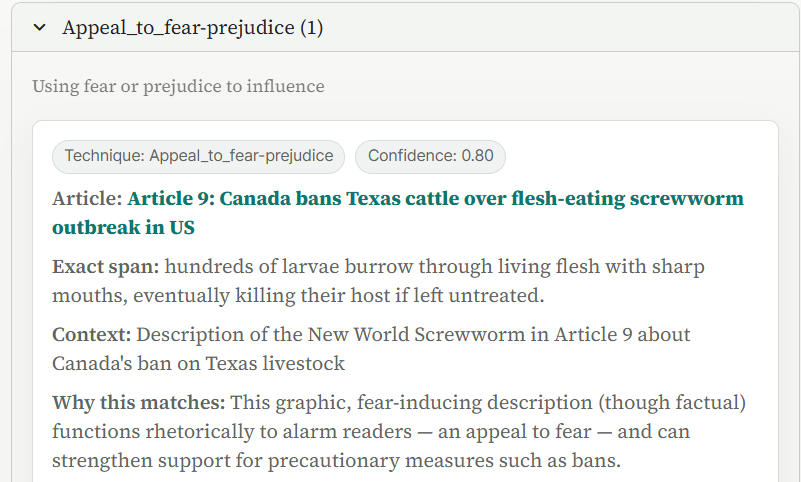}
        \caption{Example of appeal to fear}
    \end{subfigure}
    \hfill
    \begin{subfigure}{0.325\linewidth}
        \centering
        \includegraphics[width=\linewidth]{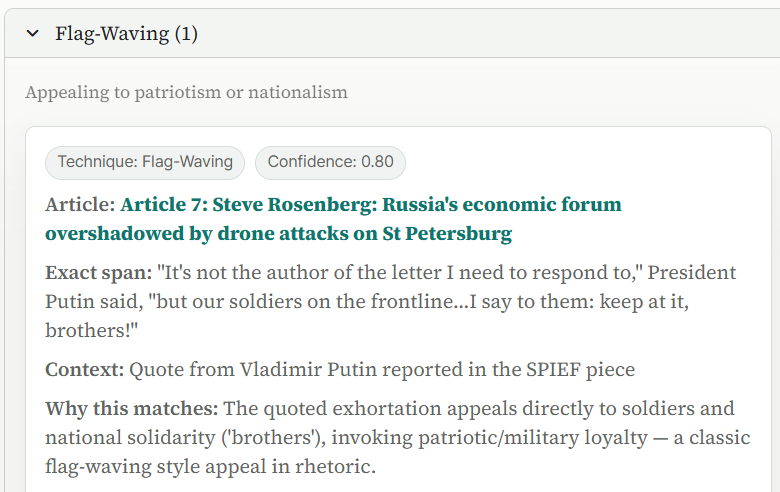}
        \caption{Example of flag waving}
    \end{subfigure}
    \hfill
    \begin{subfigure}{0.325\linewidth}
        \centering
        \includegraphics[width=\linewidth]{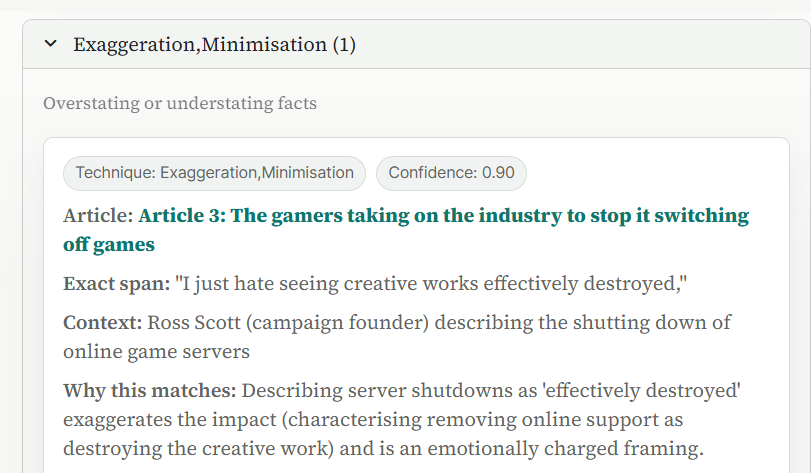}
        \caption{Example of exaggeration}
    \end{subfigure}
    \caption{Propaganda techniques detected by SCOPE in
\emph{NaturalNews.com} articles, shown with the exact triggering
spans and explanations. The system identifies Appeal to Fear/Prejudice,
Loaded Language, Causal Oversimplification, Doubt, and Appeal to
Authority --- all consistent with MBFC's Conspiracy-Pseudoscience
classification of the outlet.}
    \label{fig:propaganda_techniques_example}
\end{figure*}

\subsection{System and Components}

\begin{figure*}[t]
\centering
\includegraphics[width=0.9\textwidth]{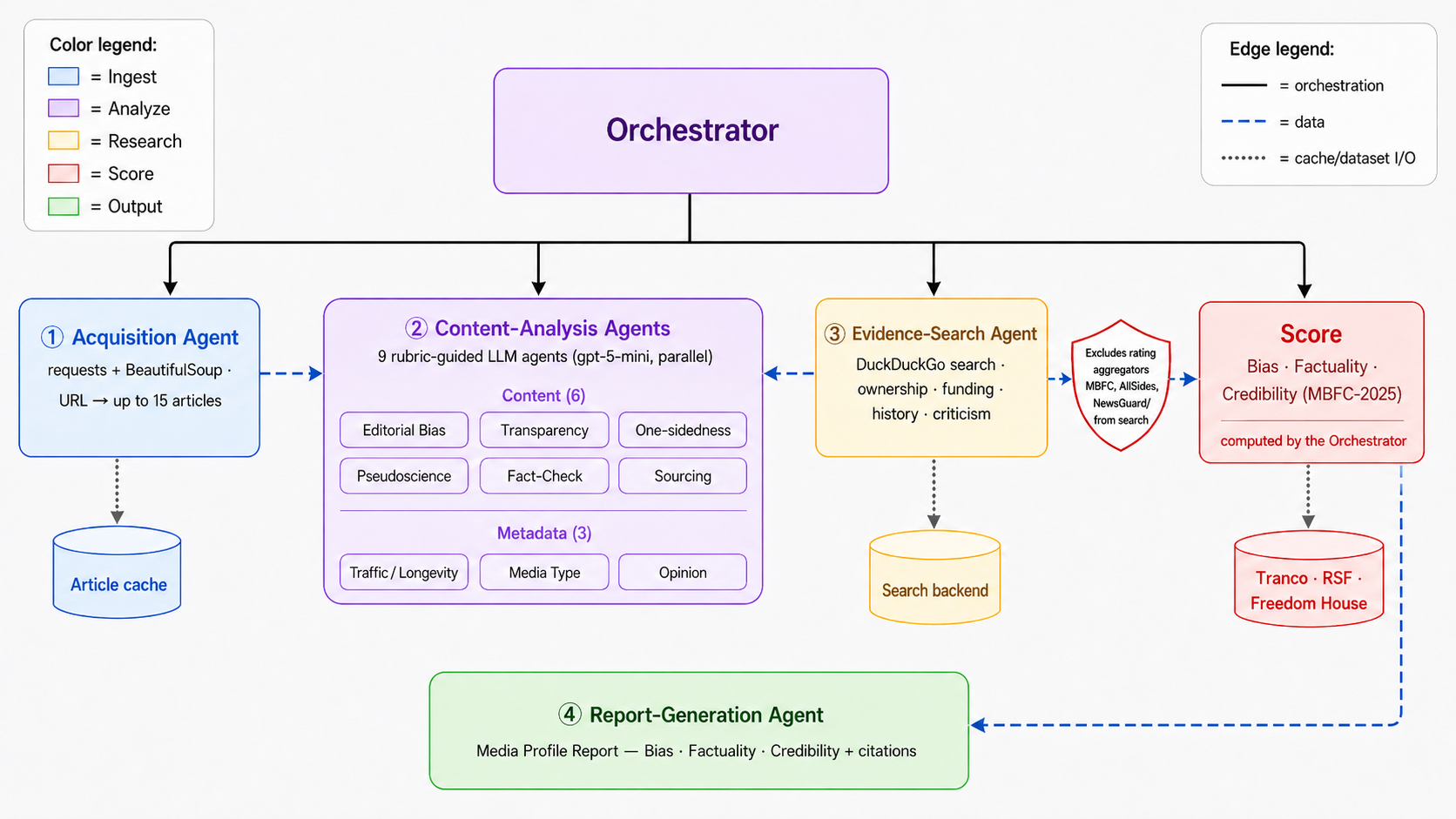}
\caption{Schematic pipeline of SCOPE system. Given an outlet URL,
four agent groups-acquisition, content analysis, evidence search, and
report generation-are coordinated by a deterministic orchestrator to
produce a structured, evidence-grounded profile.}
\label{fig:pipeline}
\end{figure*}

While MBFC relies on human reviewers to apply its methodology, our goal is to approximate this process using LLM-based agents. The evaluation procedure is coordinated by an orchestrator that manages four groups of specialised agents (Figure~\ref{fig:pipeline}). First, an \emph{acquisition} agent scrapes and caches articles from the outlet. Second, a \emph{content-analysis} group, consisting of nine agent-based analyzers, evaluates the articles according to the MBFC scoring dimensions. Third, an \emph{evidence-search} agent retrieves information about the outlet, including its history, ownership, funding, transparency ranking, and proofs of failed fact-checks. Finally, a \emph{report-generation} agent synthesises the intermediate outputs into an MBFC-style media profile.

Each agent is assigned a narrowly defined sub-task and produces a validated, schema-constrained output. The orchestrator executes the agents in a fixed order, checks whether the collected evidence is sufficient, and then computes the bias, factuality, and credibility scores. This modular design separates evidence collection, content analysis, scoring, and report generation, making intermediate decisions inspectable and the final profile auditable.

\paragraph{Scraping.}
We collect up to 15 articles per outlet using a crawler
(HTTP requests paired with BeautifulSoup parsing).
Candidate URLs are discovered from RSS feeds, Substack archives, sitemaps, and homepages, then scored by a simple keyword-based heuristic that prioritises politics, economy, and world news;
the top candidates are fetched with randomised rate-limiting to avoid blocking.

\paragraph{Analysis.}

After scraping, agents analyze editorial bias, transparency, one-sidedness, pseudoscience, fact-checking history, economic and social positioning, and sourcing quality. They examine the articles, outlet metadata, and evidence retrieved from external sources. Each agent returns a label, score, rationale, and supporting evidence.
For example, the editorial-bias agent evaluates framing, headline tone, topic selection, and the treatment of political actors. The sourcing agent examines whether factual claims cite named sources, primary documents, expert quotations, or sites. The pseudoscience agent identifies unsupported scientific, medical, and conspiratorial claims using the MBFC taxonomy.\footnote{\url{https://mediabiasfactcheck.com/pseudoscience-dictionary/}}

To estimate the economic and social components of the bias score, we use a two-axis framework. This follows the view that political ideology is not well captured by a single left--right continuum, but is better understood as having one economic and one social/cultural dimension~\citep{eysenck1954,treier-hillygus-2009,jost-etal-2009}.

The economic axis is classically defined by one's stance toward equality, redistribution, and the role of the state~\citep{bobbio1996},
while the social axis captures progressive-versus-traditional positions on cultural questions such as gender, religion, immigration, and national identity~\citep{norris-inglehart-2019}.
Both axes are scored from $-10$ to $+10$. On the economic axis, $-10$ corresponds to strongly state-interventionist positions and $+10$ to strongly free-market positions. On the social axis, $-10$ corresponds to progressive-liberal positions and $+10$ to traditional-conservative positions. Scores near zero indicate a centrist or mixed position on the corresponding axis.

For each scraped article, we first determine whether the text is relevant to either dimension. Articles that do not discuss economic policy, social values, civil liberties, or related political issues are excluded from the corresponding axis. Relevant articles are then scored on the applicable dimension. Using text to estimate ideological position follows the same general logic as work on party manifestos and media slant~\citep{budge-etal-2001,gentzkow-shapiro-2010,groseclose-milyo-2005}.
The scoring rubrics are derived from the ideology bank covering 17 policy domains. The details are given in Appendix~\ref{sec:itembank}.
We additionally check the articles for loaded language and other propaganda and manipulation techniques, consistent with \citet{da-san-martino-etal-2020-semeval} taxonomy. See examples on Figure~\ref{fig:propaganda_techniques_example}.

\paragraph{Web-Search.}
In addition to scraped articles, SCOPE uses DuckDuckGo search to collect external evidence about each outlet, including ownership, funding, editorial history, and third-party fact-checking records from organizations such as Full Fact and other independent fact-checkers.

We explicitly exclude media-rating aggregators such as MBFC, AllSides~\footnote{\url{https://www.allsides.com/unbiased-balanced-news}}, NewsGuard~\footnote{\url{https://www.newsguardtech.com/}} to prevent gold-label leakage during profiling.
 

\paragraph{External knowledge base.}
We complement outlet-derived evidence with independent reference sources. The Tranco top-1M list~\cite{lepochat-2019-tranco} estimates outlet reach using four traffic tiers: High for the top 10{,}000 domains, Medium for the top 100{,}000, Low for the top 1{,}000{,}000, and Minimal otherwise. Press-freedom ratings from the RSF Index~\cite{rsf2025} and Freedom House~\cite{freedomhouse} produce a credibility adjustment of $-1$ for limited freedom and $-2$ for no freedom.

\subsection{Demonstration interface}
\label{sec:interface}
\begin{figure*}[t]
\centering
\captionsetup[subfigure]{font=footnotesize,labelfont=bf,skip=1pt}

\begin{subfigure}[t]{0.48\textwidth}
  \centering
  \includegraphics[
    width=\linewidth,
    height=0.19\textheight,
    keepaspectratio
  ]{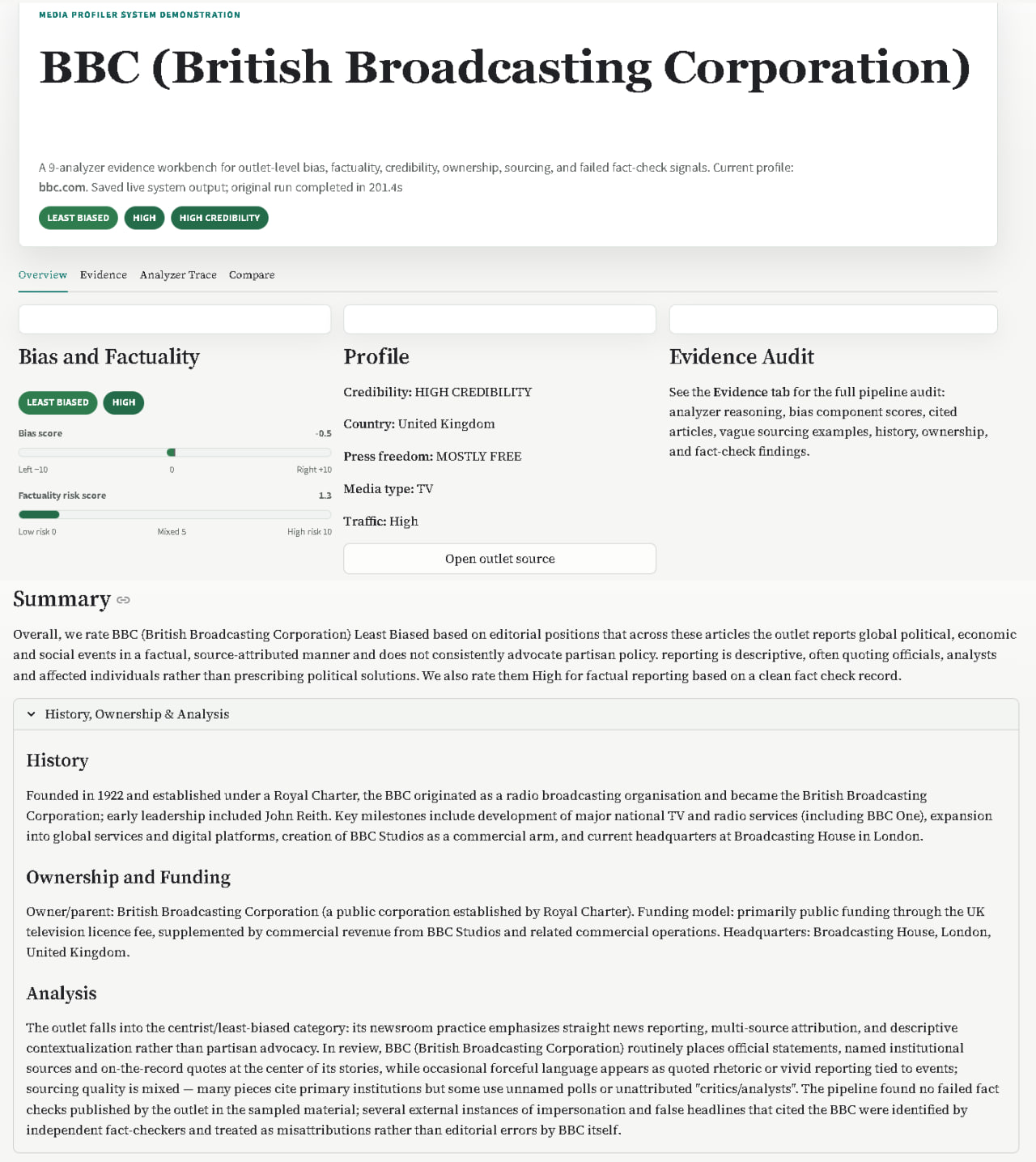}
  \caption{Overview verdict}
\end{subfigure}
\hfill
\begin{subfigure}[t]{0.48\textwidth}
  \centering
  \includegraphics[
    width=\linewidth,
    height=0.19\textheight,
    keepaspectratio
  ]{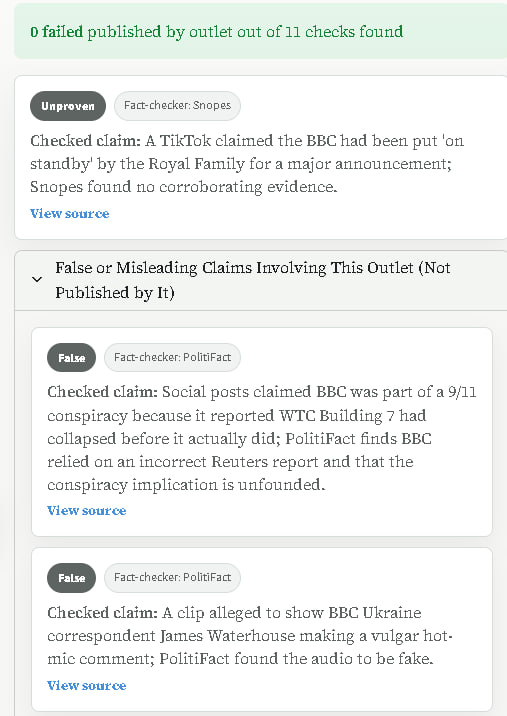}
  \caption{Failed fact-check signals}
\end{subfigure}

\vspace{0.25em}

\begin{subfigure}[t]{0.48\textwidth}
  \centering
  \includegraphics[
    width=\linewidth,
    height=0.19\textheight,
    keepaspectratio
  ]{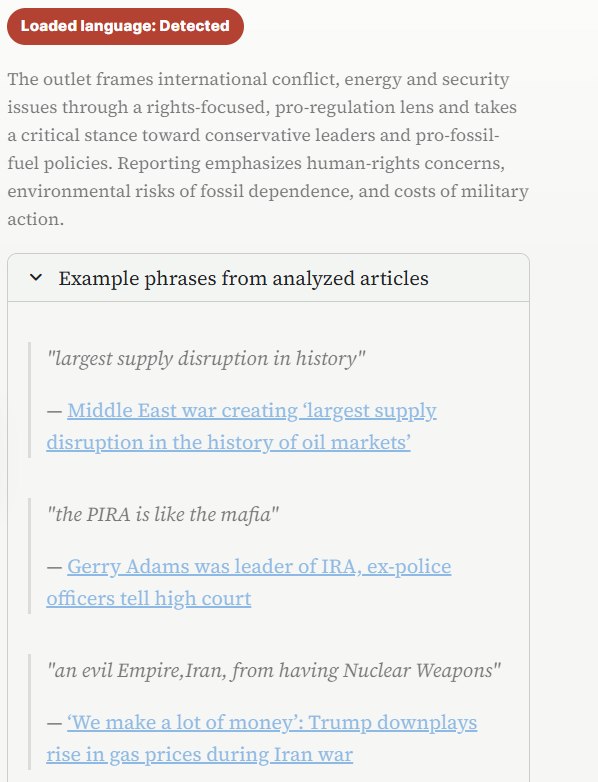}
  \caption{Analyzed-article phrases}
\end{subfigure}
\hfill
\begin{subfigure}[t]{0.48\textwidth}
  \centering
  \includegraphics[
    width=\linewidth,
    height=0.19\textheight,
    keepaspectratio
  ]{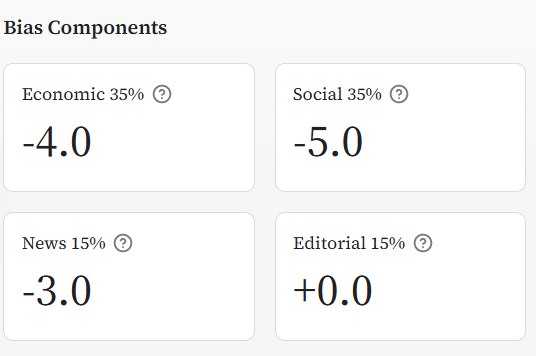}
  \caption{Bias components}
\end{subfigure}

\caption{BBC profile in SCOPE demonstration interface, showing the overview verdict, supporting failed fact-check signals, analyzed-article phrases, and bias-component scores.}
\label{fig:bbc_profile_interface}
\end{figure*}

SCOPE provides a web application that presents each outlet profile in four tabs. The \textbf{Overview} tab (Figure~\ref{fig:bbc_profile_interface}) shows MBFC-style verdicts, numeric scores, outlet metadata, and a narrative summary with collapsible history, ownership, and analysis sections. The \textbf{Evidence} tab provides an audit trail containing article quotes, analyzer findings, propaganda spans with confidence scores, and fact-check results. The \textbf{Analyzer Trace} tab displays each agent's structured output, while the \textbf{Compare} tab places two outlets side by side.

\section{Experiments and Results}
 
To demonstrate the usability of our system, we run several comparative experiments. We ablate SCOPE by comparing it against three LLM baselines. The first baseline asks the LLM to make the assessment using only its internal knowledge, without any external context. The second provides the LLM with 15 articles from the website. The third gives the LLM access to a web search tool, while excluding MBFC results to avoid label leakage.

We further compare SCOPE with Media Background Checks (MBC) by \citet{schlichtkrull-2024-generating}, a retrieval-augmented system for generating website background checks. MBC issues seven web-search queries, retrieves 30 documents per query, and uses a DeBERTa-based question-answering model~\cite{he2021deberta} to extract key facts such as founders and prior reports of misinformation.
We use the original implementation, replacing its unsupported search backend with the same search tool used in our experiments alongside the LLM to answer the questions instead of the DeBERTa-based question-answering model.

\subsection{Data}
All experiments are conducted on 200 media outlets sampled from the MBFC catalogue, selected by round-robin sampling across seven bias categories and five factuality levels for broad label coverage.

We use each outlet's MBFC bias, factuality, and credibility labels as gold annotations, and the free-text MBFC fields (summary, history, ownership, analysis) as reference text for evaluating generated reports.
All systems are evaluated on the same fixed set of collected articles.

\subsection{Setup and metrics} 
\begin{figure*}[t]
\centering
\includegraphics[width=\textwidth]{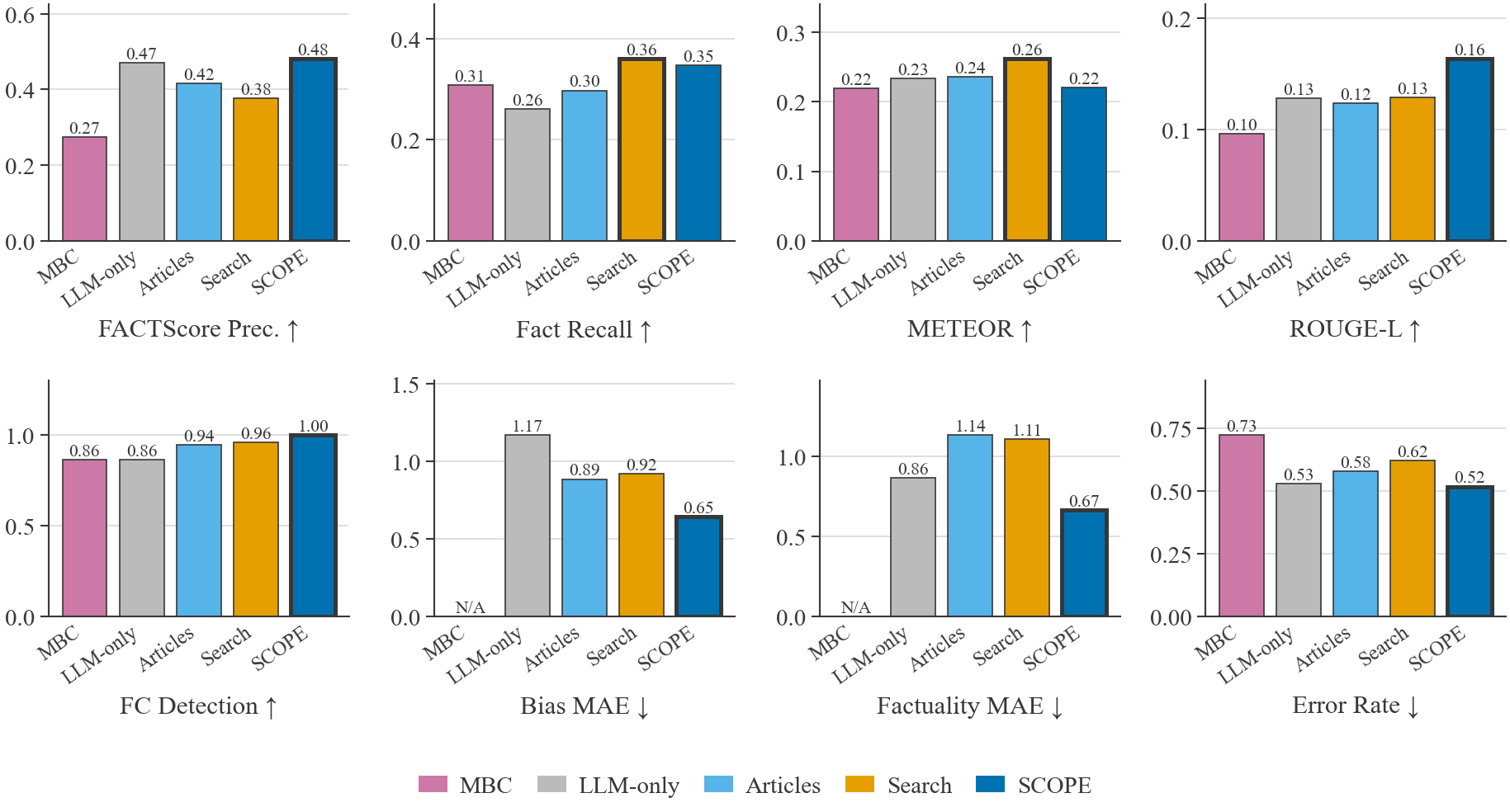}
\caption{Ablation results across eight
metrics ($\uparrow$ higher is better; $\downarrow$ lower is better).
The full \emph{SCOPE} wins on six of eight metrics, including both
ordinal MAEs, failed fact-check detection, and error rate.  \emph{Search} leads
only on recall-oriented metrics (fact recall, METEOR). We provide no Bias and Factuality evaluation for MBC, since it does not provide a direct label and gives a description only.}
\label{fig:grid}
\end{figure*}
 
 
We evaluate political leaning and factuality using MAE, following GREENER~\cite{panayotov-etal-2022-greener}, \citet{manzoor-etal-2026-multi}, and related work. MAE reflects ordinal distance: confusing Left with Left-Center is penalized less than confusing it with Right. Bias is ordered from Extreme-Left, Left, Left-Center, Least Biased, Right-Center, Right, to Extreme-Right; factuality ranges from Very Low to Very High through Low, Mixed, and High. We assess reports with FACTScore~\cite{min-etal-2023-factscore}: precision measures claim support, recall measures MBFC-reference coverage, and error rate equals one minus precision.
FC-Detection measures how often the system finds an outlet's failed fact checks. We count only checks of content published by the outlet, not checks about the outlet itself.

To assess the content overlap of the generated reports, we also compute METEOR~\cite{banerjee-lavie-2005-meteor} and ROUGE-L~\cite{lin-2004-rouge}, using the MBFC text as the reference. All experiments use GPT-5 mini, the Pareto-optimal model in the cost--quality comparison of Appendix~\ref{sec:model-choice}.

\subsection{Main results}
\label{sec:ordinal}
Figure~\ref{fig:grid} summarises the comparison across the approaches.

SCOPE is the strongest overall, achieving the lowest bias MAE ($0.65$ vs.\
$1.17$ for the LLM-only baseline), the lowest factuality MAE ($0.67$),
perfect fact-check detection ($1.00$), the best ROUGE-L ($0.16$), the
highest FACTScore precision ($0.48$), and the lowest error rate ($0.52$).
 

It also always finds if there are failed fact-checks by the outlet (as, for example, for \emph{NaturalNews.com}, as shown in the Appendix~\ref{sec:factcheck-example})

Equipping the LLM with a web-search tool
raises fact recall ($0.36$) and METEOR ($0.26$), but also produces the
highest error rate ($0.62$), showing that live web text increases coverage
at the cost of unsupported assertions. Adding offline articles improves
bias calibration over the LLM-only baseline ($0.89$ vs.\ $1.17$) but
slightly hurts factuality, consistent with article text being more
informative about an outlet's stance than about its factual track record.

On the ordinal axes, SCOPE predicts the exact bias class for $50\%$ of
outlets and lands within one ordinal class for $90\%$; on factuality it is
exact for $42\%$ and within one class for $90\%$ (Figure~\ref{fig:ordinal};
full breakdown in Appendix~\ref{sec:ordinal-acc}).

The full pipeline costs only \$0.054 per outlet---comparable to the LLM-only baseline (\$0.050) and roughly half the cost of Search (\$0.101).


 


 

\section{Conclusion}
We presented SCOPE, a multi-agent system that transforms a single outlet
URL into a structured, evidence-grounded profile of political bias,
factual reliability, and credibility.
SCOPE divides media profiling among specialised agents for acquisition, content analysis, evidence search, and report generation. Their typed, schema-constrained outputs are visible in the interface, making each verdict traceable to its evidence rather than an opaque label.

On a 200-outlet MBFC benchmark, SCOPE achieves the lowest bias and factuality MAE and perfect failed-fact-check detection among the evaluated baselines. It also keeps $90\%$ of predictions within one ordinal class on both axes, at about \$0.05 per outlet.

 
Qualitative analysis shows that SCOPE's evidence remains useful even when its final label is incorrect, especially for conspiracy and pseudoscience outlets that do not fit a left--right scale. We release the system, live demo, and leakage-controlled evaluation code.
 
In future work we aim
to extend coverage beyond English-language and U.S.-centric outlets, and
to generalise the agent scaffolding to adjacent entities such as
politicians, pollsters, and government bodies.

\section*{Limitations}
Although SCOPE performs well in our evaluation, several limitations remain and should be considered when interpreting the results.

First, our evaluation focuses on U.S.-centric, English-language outlets because labeled fact-checking and media-profiling data are scarce for other languages and regions. SCOPE itself is not limited to English and can work with other languages. However, the lack of suitable labels prevented us from evaluating its multilingual performance. The reported results therefore may not fully generalize to non-Western, non-English, or low-visibility outlets.

Second, while we explicitly filter web-search results to exclude MBFC, AllSides, NewsGuard, and similar media-rating platforms, indirect leakage is still possible, since other sites may reproduce or cite their ratings. We manually inspected the retrieved web-search traces in our experiments and did not find evidence of such leakage, but this risk cannot be ruled out completely.

Third, the sourcing analyzer may miss hyperlink metadata. The scraper converts each article to plain text, retaining the displayed anchor text but removing the target URL and any indication that the text was clickable. A citation embedded behind a phrase may therefore be invisible to the analyzer unless the source name or URL also appears in the text. In manual testing, this sometimes led the system to report missing evidence even when the article linked to a relevant source. We will address this in the next release.

Finally, conspiracy and pseudoscience outlets can be difficult to place on a left--right political scale. SCOPE usually captures their low factuality and credibility, but it may occasionally assign the wrong political direction when the outlet's content is anti-establishment, conspiratorial, or ideologically mixed.

\section*{Ethics and Broader Impact}
SCOPE scores politically sensitive attributes of news outlets, and such
scores can be misused to dismiss or amplify sources. Every score is backed by inspectable intermediate evidence and withheld when the evidence-sufficiency gate is not met. The system is intended as an auditable assistive tool for researchers, journalists, and analysts---not as an arbiter of truth.

\bibliography{acl_latex_clean_version.bib}

\appendix

\section{Model Selection}
\label{sec:model-choice}
\label{sec:appendix}

We selected the model by evaluating five candidates on the same set of outlets ($\approx$50), while keeping the token cap fixed, so that differences in quality-per-dollar reflect model pricing rather than usage variation.

Quality was measured as an equal-weight, min--max-normalised composite of the task metrics. We divided this quality score by cost to obtain the quality-per-dollar metric reported in Table~\ref{tab:pareto}.

By this metric, GPT-5 mini is Pareto-optimal: none of the evaluated models is both cheaper and higher quality. It achieves a score of 12.1, about $3\times$ higher than premium models such as GPT-5 and GPT-5.2. This comparison is based on the LLM-only setting.

\begin{center}
\scriptsize
\setlength{\tabcolsep}{3pt}
\renewcommand{\arraystretch}{0.8}
\resizebox{\columnwidth}{!}{%
\begin{tabular}{@{}l r r r@{}}
\toprule
\textbf{Model} & \textbf{\$/outlet} & \textbf{Quality} & \textbf{Quality/\$} \\
\midrule
GPT-5 nano          & 0.0110 & 0.086 & 7.8  \\
\textbf{GPT-5 mini} & \textbf{0.0507} & \textbf{0.614} & \textbf{12.1} \\
GPT-4.1             & 0.2201 & 0.858 & 3.9  \\
GPT-5               & 0.2491 & 0.980 & 3.9  \\
GPT-5.2             & 0.3483 & 0.576 & 1.7  \\
\bottomrule
\end{tabular}}
\captionsetup{hypcap=false}
\captionof{table}{Cost--quality trade-off in LLM-only mode.
Quality is an equal-weight, min--max-normalised composite over five task
metrics. The recommended operating point is shown in \textbf{bold}.}
\label{tab:pareto}
\end{center}

\section{Ordinal Accuracy}
\label{sec:ordinal-acc}
\begin{figure}[ht]
\centering
\includegraphics[width=0.92\columnwidth]{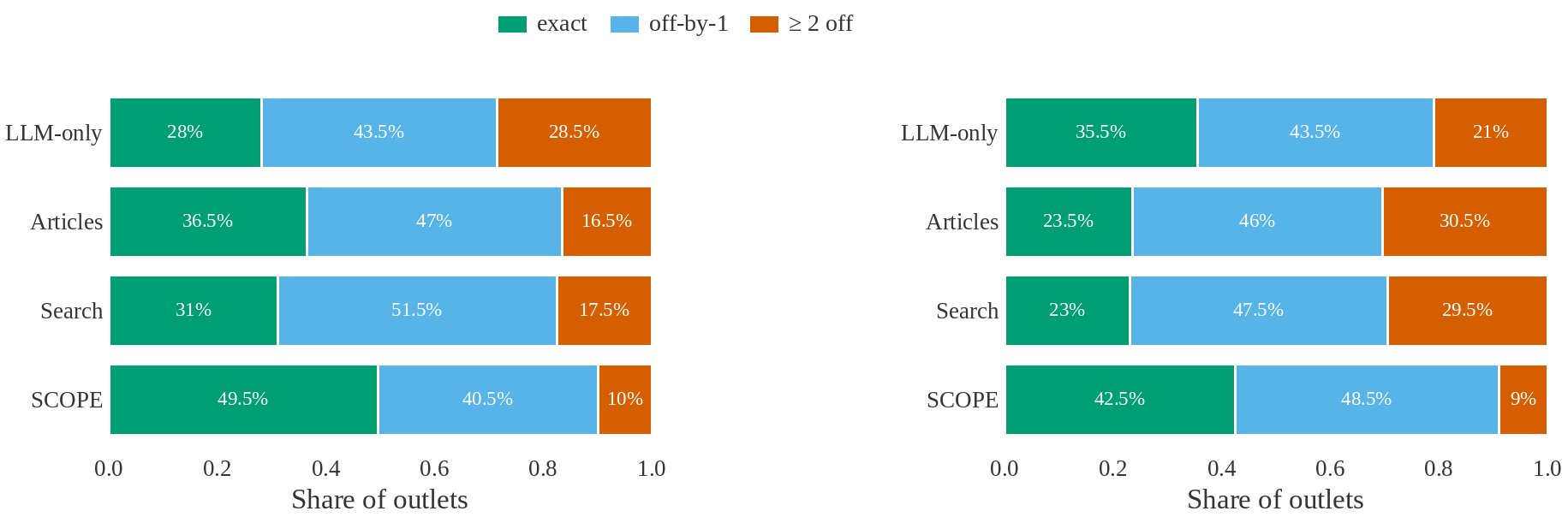}
\caption{Ordinal accuracy. It shows the share of outlets
whose predicted class is exact, off by one, or ${\geq}2$ classes from the
gold label, for bias (left) and factuality (right). }
\label{fig:ordinal}
\end{figure}

Figure~\ref{fig:ordinal} gives the per-class breakdown, for both the
bias and factuality axes, behind the aggregate within-one-class figures
reported in Section~\ref{sec:ordinal}.

It shows that \emph{SCOPE} keeps roughly $90\%$ of predictions within one class for both factuality and bias.

\section{Failed Fact-Check Detection Example}
\label{sec:factcheck-example}
\begin{figure}[ht]
\centering
\includegraphics[width=0.82\columnwidth]{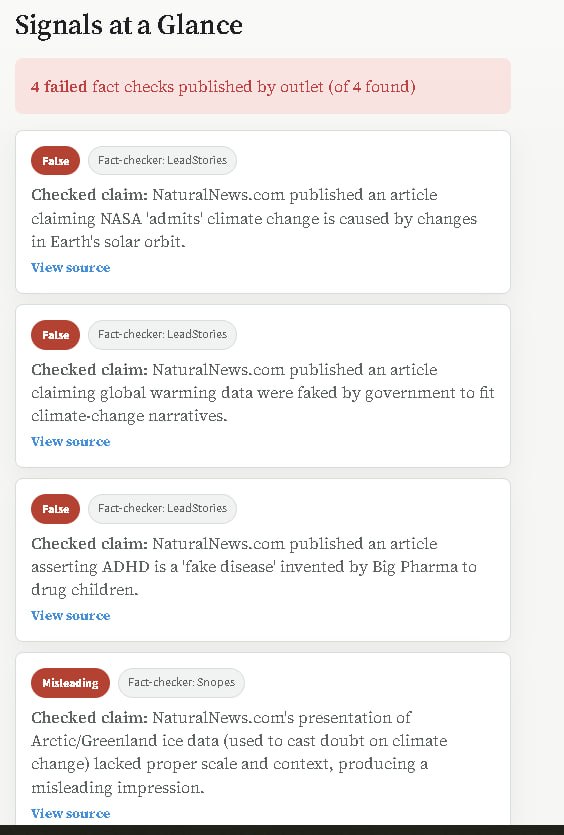}
\caption{``Signals at a Glance'' panel for NaturalNews.com: SCOPE
locates four failed fact-checks published by external fact-checkers
(LeadStories, Snopes).}
\label{fig:nn-factchecks}
\end{figure}

Figure~\ref{fig:nn-factchecks} shows the ``Signals at a Glance'' panel
SCOPE produces for \emph{NaturalNews.com}.

\section{The Ideology Item Bank}
\label{sec:itembank}
\begin{table*}[t]
\centering
\footnotesize
\setlength{\tabcolsep}{4pt}
\renewcommand{\arraystretch}{1.05}
\begin{tabular}{@{}p{0.225\textwidth} p{0.355\textwidth} p{0.355\textwidth}@{}}
\toprule
\textbf{Policy domain} & \textbf{Left / progressive pole (grounding)} & \textbf{Right / conservative pole (grounding)} \\
\midrule
\multicolumn{3}{@{}l}{\emph{Economic axis}\quad($-10$ state-interventionist $\rightarrow +10$ free-market)} \\
\midrule
Government ownership \& nationalisation & Public ownership of the means of production \citep{marx-engels-1848} & Private enterprise, the ``invisible hand'' \citep{smith1776} \\
Taxation \& redistribution & Progressive and wealth taxation \citep{piketty2014,saez-zucman-2019} & Low, flat taxation; limited state \citep{friedman1962,hayek1944} \\
Labour, unions \& minimum wage & Collective bargaining, wage floors \citep{freeman-medoff-1984,card-krueger-1995} & Market-set wages, labour deregulation \citep{friedman1962} \\
Corporate regulation \& antitrust & Aggressive antitrust enforcement \citep{wu-2018} & Deregulation; regulatory-capture critique \citep{stigler-1971} \\
Trade \& globalisation & Managed / fair trade, worker protection \citep{rodrik-2011} & Free trade, comparative advantage \citep{ricardo1817} \\
Social-welfare programmes & Social-democratic welfare state \citep{esping-andersen-1990} & Welfare-dependency critique \citep{murray-1984} \\
Financial-sector regulation & Post-crisis bank regulation \citep{stiglitz-2010} & Market discipline, free banking \citep{friedman1962} \\
\midrule
\multicolumn{3}{@{}l}{\emph{Social axis}\quad($-10$ progressive-liberal $\rightarrow +10$ traditional-conservative)} \\
\midrule
LGBTQ+ rights & Equal rights, anti-discrimination \citep{butler1990} & Traditional family / religious objection \citep{burke1790} \\
Abortion \& reproductive rights & Bodily autonomy as individual liberty \citep{mill1859} & Traditional morality, foetal personhood \citep{burke1790} \\
Immigration (social / cultural) & Permissive borders, inclusion \citep{carens2013,hainmueller-hopkins-2014} & Restriction; cultural-cohesion priority \citep{norris-inglehart-2019} \\
Climate change \& environment & Urgent, human-caused crisis \citep{mccright-dunlap-2011} & Cost-sceptic, anti-regulation \citep{lomborg-2001} \\
Racial justice \& equity & Anti-racism, structural remedy \citep{kendi2019,coates-2014} & Colour-blind individualism \citep{sowell1987} \\
Gender equality & Feminist / structural equality \citep{butler1990} & Equity-feminism critique \citep{sommers-1994} \\
Gun rights \& control & Stronger firearm regulation \citep{jost-etal-2009} & Robust individual right \citep{sowell1987} \\
Criminal justice & Decarceration, reform / abolition \citep{davis-2003} & Deterrence, tough-on-crime \citep{sowell1987} \\
Religion \& secularism & Strict church--state separation \citep{mill1859} & Religion in public life \citep{burke1790} \\
Education policy & Universal public provision \citep{rawls1971} & School choice and vouchers \citep{friedman-1955} \\
\bottomrule
\end{tabular}
\caption{The 17 policy domains of the ideology item bank, with a
representative source grounding each pole. Every domain expands into a
4-left / 4-right / 1-centre ladder of observable editorial stances, each mapped to a calibrated point on the
$-10$ to $+10$ economic or social axis and individually cited in the
released bank. These two scales and the 17 domains are distilled into the
single scoring prompt described in the main text; the items themselves are
not executed per outlet.}
\label{tab:itembank}
\end{table*}

The economic and social bias axes are anchored by an offline ideology item bank.

The bank contains reference-backed indicators organised into 17 policy domains: 7 economic domains and 10 social domains.

Each domain contains nine possible positions: four left/progressive stances, one centrist stance, and four right/conservative stances.

The bank's purpose is to make the scoring scale explicit: each item describes an observable editorial stance and maps it to a point on the corresponding $-10$ to $+10$ axis, and in the released version is annotated with a primary or scholarly source supporting that mapping.

Table~\ref{tab:itembank} reports the 17 domains with a representative source for each pole; where no single canonical source defines a pole, we instead cite literature showing that the issue loads on the economic or social/cultural dimension of ideology~\citep{jost-etal-2009,treier-hillygus-2009,budge-etal-2001}.

Concretely, the prompt embeds the two axis scales and domain summaries as reference points, fixing the scale's meaning before the model assigns each outlet's economic and social scores in one step.

\end{document}